\documentclass[proof]{WileyASNA-v1}

\articletype{Article Type}%

\received{26 April 2016}
\revised{6 June 2016}
\accepted{6 June 2016}

\begin{document}

\title{The Southern Wide-Field Gamma-ray Observatory (SWGO)\protect\thanks{\url{https://www.swgo.org}}}

\author[1]{Ulisses Barres de Almeida*}

\author[2]{on behalf of the SWGO Collaboration}


\authormark{U. BARRES DE ALMEIDA \textsc{et al}}

\address[1]{\orgname{Brazilian Center for Physics Research (CBPF)}, \orgaddress{\state{Rio de Janeiro}, \country{Brazil}}}

\corres{*Ulisses Barres de Almeida, \email{ulisses@cbpf.br}}

\presentaddress{Brazilian Center for Physics Research (CBPF), Rua Dr. Xavier Sigaud 150, 22290-180 Rio de Janeiro, Brazil.}

\abstract{The scientific potential of a wide field-of-view, and very-high duty cycle, ground-based gamma-ray detector has been demonstrated by the current generation of instruments, such as HAWC and ARGO, and will be further extended in the Northern Hemisphere by LHAASO. Nevertheless, no such instrument exists in the Southern Hemisphere yet, where a great potential lies uncovered for the mapping of Galactic large scale emission as well as providing access to the full sky for transient and variable multi-wavelength and multi-messenger phenomena. Access to the Galactic Centre and complementarity with the CTA-South are other key motivations for such a gamma-ray observatory in the South. There is also significant potential for cosmic ray studies, including investigation of cosmic-ray anisotropy. In this contribution I will present the motivations and the concept of the future Southern Wide-Field Gamma-ray Observatory (SWGO), now formally established as an international Collaboration and currently in R\&D phase. I will also outline its scientific objectives.}


\jnlcitation{\cname{%
\author{Barres de Almeida U.}, 
\author{The SWGO Collaboration}} (\cyear{2020}), 
\ctitle{The Southern Wide-Field Gamma-ray Observatory (SWGO)}, Proc. of IWARA 2020, \cjournal{AN}, \cvol{2021;??:?--?}.}

\fundingInfo{The author acknowledges the support of a CNPq Productivity Research Grant no. 311997/2019-8 and a Serrapilheira Institute Grant no. Serra - 1812-26906. He also acknowledges the receipt of a FAPERJ Young Scientist Fellowship no. E-26/202.818/2019.}

\maketitle


\section{Observational Panorama of Gamma-ray Astronomy}\label{sec1}

Direct detection of primary astrophysical gamma-rays is only possible with satellite-based detectors, such as the Fermi Telescope\footnote{https://fermi.gsfc.nasa.gov}, but the cost and size restrictions associated to space instrumentation limit their collection area and sensitivity. As fluxes become too small towards higher energies -- typically above 100 GeV -- satellites are no longer an option for astronomical observations. In this case, one needs to recourse to the showers of particles created as a result of the gamma-ray interaction with the atmosphere for indirect detection. 

The atmospheric showers can be studied with observatories of two complementary types: imaging atmospheric Cherenkov telescopes (IACTs) and extensive air-shower arrays, EASs (Fig.1). The Cherenkov Telescopes such as CTA, are highly sensitive pointing instruments, but with a limited duty-cycle and narrow field-of-view. Wide-field observatories, such as the proposed SWGO, are instead ideal to study the emission from very extended regions of the sky and to search for transient sources, serving as a complement to the IACTs by providing alerts for follow-ups. The air-shower arrays also have the highest energy reach of any gamma-ray observatory.

The first very-high-energy (VHE) gamma-ray emission from an astrophysical source was observed 30 years ago from the Crab Nebula~\citep{1989ApJ...342..379W}. Since then, hundreds of sources have been discovered at these extreme energies, both from Galactic and extragalactic origins.\footnote{http://tevcat2.uchicago.edu} Many such sources present variability, with flare durations that range from days to minutes, or even seconds in the case of the Gamma-ray Bursts (GRBs). Such transient phenomena are at the heart of multi-messenger astrophysics, being potential electromagnetic counterparts to neutrino~\citep{2019FrASS...6...32H} and gravitational wave detections\footnote{See \url{https://kilonova.space} for a multi-messenger open database associated to gravitational wave events}, and their study greatly benefits from instruments which are able to continuously monitor large portions of the sky at energies reaching well-above those attainable by satellite-based experiments.

\begin{figure*}[ht!]
\centerline{\includegraphics[width=342pt,height=25pc]{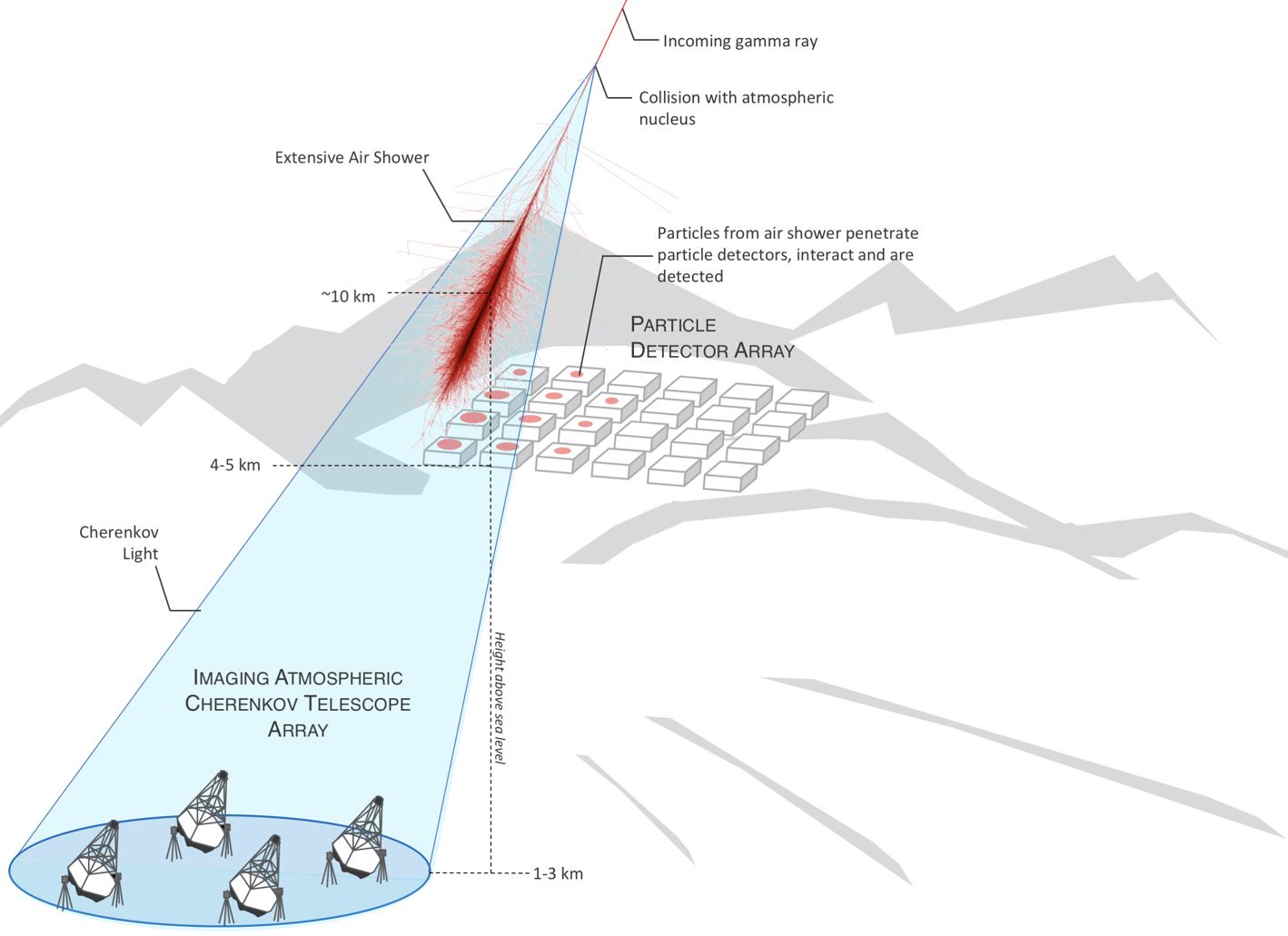}}
\caption{Illustration of the complementary ground-based detection techniques of high-energy gamma rays. Credits: Richard White, MPIK, and the SWGO Collaboration.\label{fig1}}
\end{figure*}

\section{The Highest-Energy Frontier}\label{sec2}

The third generation of IACTs (represented by H.E.S.S., MAGIC and VERITAS) has shown the VHE gamma-ray sky to be richly populated with powerful particle accelerators. The over 200 astrophysical objects known to emit above 100 GeV pertain to a wide variety of source classes, including supermassive black holes, compact galactic objects and their associated environments, starburst galaxies and binary systems, such as micro-quasars and X-ray binaries. About 30\% of the VHE objects remain  unassociated with sources in other bands. 

The deepest and most detailed systematic view of the VHE sky to date is that provided by the H.E.S.S. Galactic Plane Scan (HGPS). Conducted within latitudes $|b| < 3^{\circ}$ and a longitude range between 65$^{\circ}$ and 250$^{\circ}$, it is complete down to a point-source sensitivity of circa 15 mCrab~\citep{2018A&A...612A...1H}. 

At the highest energies, above several 10s TeV, our most complete view of the sky comes from the EAS technique. The Third HAWC Catalogue (3HWC) contains 65 sources detected after 1523 days of wide-field exposure~\citep{2020arXiv200708582A}. Given HAWC's opposite geographical location to H.E.S.S., both surveys are mostly complementary in terms of their sky coverage, and overlap only in a small portion of the Galactic Plane. Although a potential counterpart has been identified in the Fermi-LAT catalogue for several of the non-overlapping sources in the 3HWC, a southern-hemisphere EAS observatory is much awaited, should we wish to disclose a fuller picture of the population of Galactic accelerators, specially when CTA enters operation.

Particularly noticeable at these highest energies was the detection of extended gamma-ray emission from the direction of old pulsars ($>$100 kyr of age), which have broken out from their SNR shells. The degree-wide VHE emission seem from such sources matches the predicted TeV halos originating from the inverse-Compton scattering of the CMB by relativistic electrons diffusing from the pulsar. Two of the pulsars listed in the 3HWC, PSR J0631+1036 and PSR J1740+1000, have not been previously detected at TeV energies, for example, and give a hint of the potential for new discoveries in field.

\begin{figure}[h!]
	\centerline{\includegraphics[width=76mm,height=11pc]{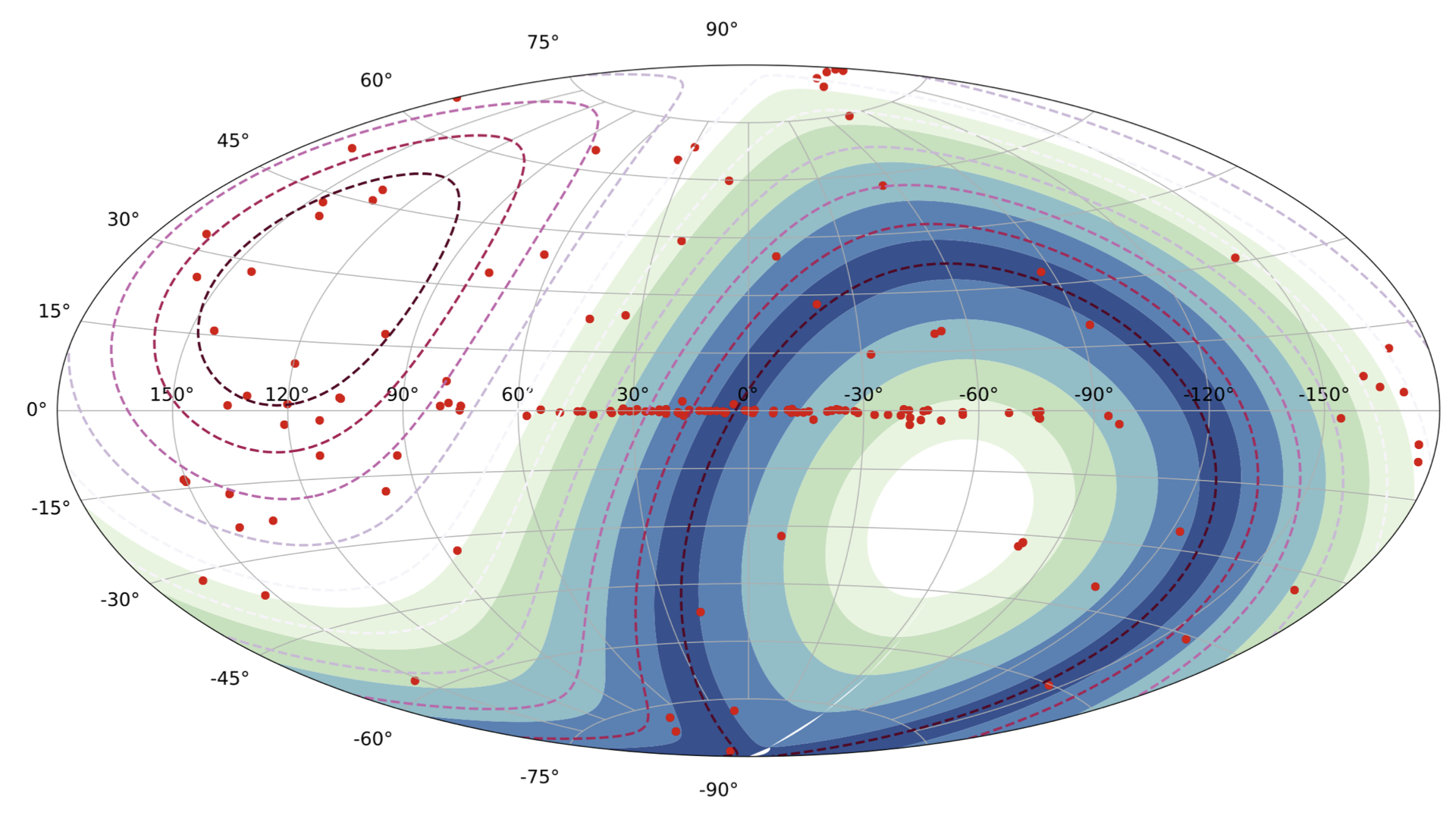}}
	\caption{The colour gradient shows the $\pm 45^{\circ}$ zenith visibility for a wide-field instrument located at $25^{\circ}$ latitude south. Red dots mark the positions of known VHE sources. SWGO will complement the view of the Galactic source population towards the highest energies. Credits: SWGO Collaboration. \label{fig2}}
\end{figure}

Detailed spectral studies of the sources presented in the 3HWC are still missing, but already in the 2nd HAWC Catalogue~\citep{2017ApJ...843...40A} 4 sources -- namely MGRO 2019+371, MGRO 1908+06, HESS J1825+137, Crab Nebula -- had been detected above 100 TeV, an energy frontier that starts to approach the PeV particle acceleration target, crucial for probing the origin of Galactic Cosmic-Rays. In the particular case of the Crab Nebula, its log-parabolic spectrum has been seen to extend unabated beyond 100 TeV by the Tibet AS array ~\citep{2019PhRvL.123e1101A}.

Concerning extragalactic sources, the latest high-energy window opened comes from the detection, respecively by MAGIC and H.E.S.S., of sub-TeV emission from the early~\citep{2019Natur.575..455M} and late afterglow phases~\citep{2019Natur.575..464A} of two gamma-ray bursts (GRBs). These detections were a long-sough goal of the field, and revealed unexpected spectral features at the VHE band, indicating the presence of a previously unknown, long-lasting inverse-Compton component to the afterglow emission~\citep{2019Natur.575..459M}. Such results are promising for the perspectives of wide-field instruments for detecting GRBs, as these should be able to provide serendipitous alerts and directly probe the onset of the VHE emission or even the GRB prompt phase.

\section{A Wide-Field Observatory in the South}\label{sec3}

The SWGO Collaboration was founded in July 2019 by a group of about 40 institutions from 9 countries as an international R\&D Project to develop what would be the first extensive air-shower array for gamma-ray astronomy in the Southern Hemisphere.\footnote{www.swgo.org} The new observatory is aimed as a southern version of the current ground-based detector arrays HAWC, in Mexico, and LHAASO, in China, on which the project is benchmarked. The baseline design constitutes of water Cherenkov detector unities to sample the gamma ray-induced particle showers in the atmosphere, by measuring the light produced when the shower particles cross the water tanks.

\subsection{The core concept of the observatory}

SWGO plans to explore new layouts and technologies in order to increase the overall sensitivity as well as the low energy detection threshold and its highest energy reach. It also aims at enabling each WCD unity with a muon tagging capability in order to enhance gamma-hadron separation and open up the possibility for mass-resolved cosmic-ray anisotropy and composition studies in the energy range up to PeV.

The new observatory is planned to be installed in the Andes, at or above 4.4 km a.s.l., as the high-altitude frontier is crucial for our plans of pushing the energy threshold towards the 100 GeV range (see e.g.,~\cite{2009NJPh...11e5007S}). Such energy window remains closed as far as wide-area ground-based detectors are concerned, and store a unique scientific potential for studying the most extreme VHE transients such as GRBs. At the highest energy range the goal is to reach a good sensitivity to gamma-rays beyond 100 TeV, in order to detect PeVatrons. 

The current baseline reference detector configuration is generally based on~\cite{2017ICRC...35..819S} and~\cite{2018APh....99...34A}, and consists in the combination of a high fill-factor core of circa 80,000 m$^2$ with instrumented area above 80\% (equivalent to 5x the HAWC instrumented area), and a significantly larger and sparse outer array of at least 200,000 m$^2$, to provide at one time a low energy detection threshold and good sensitivity to high-energy events. Currently, multiple WCD detector unities are under consideration, including solutions based on tanks (HAWC-like), ponds (LHAASO-like) or for installation directly in an natural lake. The final choice is likely to depend strongly on the site choice, as well as considerations such as water access, feasibility and construction costs, and compatibility with science-driven design goals. The goal for the R\&D study is to finalise a design proposal in 2022. 

\begin{boxtext}
\section*{The Observatory Core Concept}

\noindent \textbf{I -} High-altitude particle detector above 4.4 km a.s.l;\\

\noindent \textbf{II -} Latitude range between 15$^{\circ}$ and 30$^{\circ}$ latitude south;\\

\noindent \textbf{III -} Wide energy range reaching down to 100 GeV and beyond 100 TeV;\\

\noindent \textbf{IV -} High fill-factor core (> 4x HAWC) for significantly better (> 10x) sensitivity, plus large, sparse outer array;\\

\noindent \textbf{V -} WCD units with  muon counting capability.\\
\end{boxtext}

\begin{figure*}[ht]
	\centerline{\includegraphics[width=440pt,height=20pc]{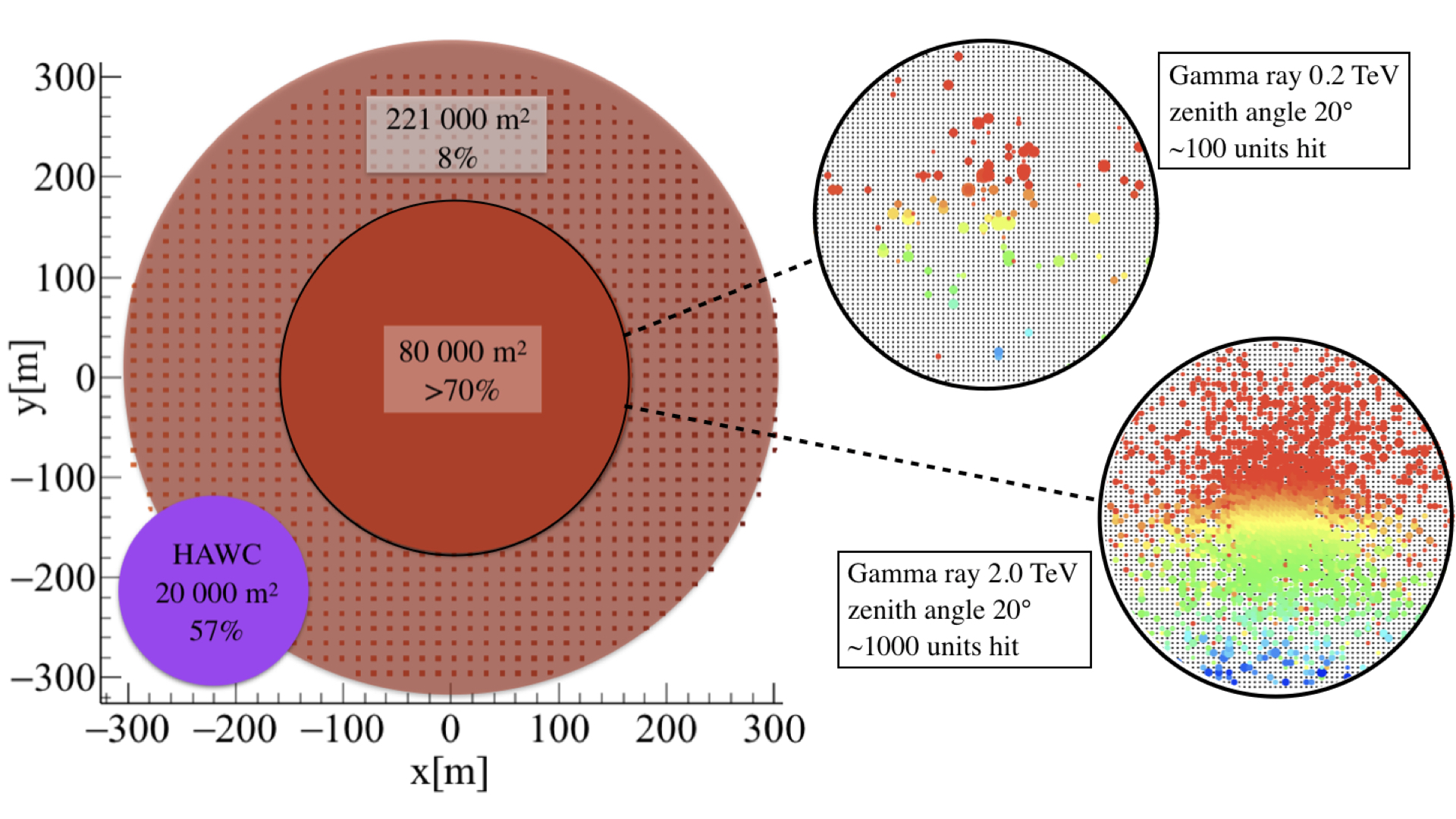}}
	\caption{The SWGO "Strawman" reference detector layout, as outlined in~\cite{2017ICRC...35..819S}. The left image schematically shows the baseline configuration, with a high fill-factor inner core and a sparse outer array of WCD units. In the right, the signals of a 200 GeV and 2 TeV gamma-ray simulated showers are shown, evidencing the importance of the dense core for the detection of low-energy events. \label{fig3}}
\end{figure*}

\section{The Core Science Case}\label{sec4}

We now briefly outline the science case for SWGO, for which the core programme and the principal science benchmarks are being currently developed to guide the design goals of the Observatory. As shown in Table 1, the core science case for SWGO in the VHE energy range between 0.1-100 TeV rests on four main pillars: 

\begin{itemize}
\item \textbf{short-timescale phenomena}, including the study of gamma-ray bursts (GRBs) and the detection of AGN transient events~\citep{2020MNRAS.497.3142L};
\item \textbf{galactic accelerators}, in particular the search for PeVatrons, the study of gamma-ray halos and pulsar-wind nebulae~\citep{2020A&A...636A.113G}, as well as measuring the diffuse gamma-ray emission and Fermi Bubbles at VHE; 
\item \textbf{dark matter}, for which SWGO would have the potential to probe the entire WIMP mass-range, up to 100 TeV~\citep{2019JCAP...12..061V}.
\item \textbf{cosmic ray} composition and anisotropy studies~\citep{2019ApJ...871...96A}.  
\end{itemize}

Below we will develop further each of the science cases. A more thorough discussion can be found at~\cite{2019arXiv190208429A}.

\paragraph{Galactic Accelerators}
The study of the Galaxy in VHE gamma-rays, including the GC and the Fermi Bubbles, is fundamental to search for particle accelerators and probe the origin of Galactic CRs. In fact, the direct detection of PeVatrons, whose Galactic population remains largely unknown~\citep{2019arXiv190908609H}, constitutes one of the fundamental science goals of SWGO. 

The direct gamma-ray signatures of PeV protons are generally expected to appear at or above 100 TeV. Our general lack of knowledge concerning the behaviour of these potential PeVatron accelerators (i.e., presence of cut-offs in the power law spectrum; e.g., ~\cite{2018MNRAS.479.3415C}) set strong requirements both on the sensitivity and energy resolution of the observatory at this extreme energy range (see Table 1), should we be able to probe a significant fraction of e.g. the CTA Galactic source ctalogue, for signatures of PeV particle population (e.g.,~\cite{2019ICRC...36..618A}).

Another fundamental frontier to be probed by SWGO is related to extended sources, and in particular gamma-ray halos associated to old PWNe~\citep{2020A&A...636A.113G}. Here SWGO is well placed to complement CTA in studying these sources at energies beyond 10 TeV and extend the sensitivity reach to objects that are a few degrees in angular extent. Although source confusion is expected to be significant for these most extended sources, SWGO should still be able to resolve most of the PWNe in the HGPS~\citep{2018A&A...612A...2H}.

\begin{center}
\begin{table*}[t]%
\caption{SWGO Core Science Goals and Associated Preliminary Design Considerations\label{tab1}}
\centering
\begin{tabular*}{500pt}{@{\extracolsep\fill}lccD{.}{.}{3}c@{\extracolsep\fill}}
\toprule
&\multicolumn{1}{@{}c@{}}{\textbf{Scientific Targets\tnote{$^1$}}} & \multicolumn{1}{@{}c@{}}{\textbf{Performance Goals\tnote{$^2$}}}\\ 
\textbf{Core Science Case} & \textbf{}  & \textbf{}  & \multicolumn{1}{@{}l@{}}{\textbf{}}  & \textbf{}   \\
\midrule
Short-timescale Phenomena & GRB detection and spectral measure  & $F_{<0.3\rm{TeV}}\sim10^{-11}$~erg/cm$^{2}.\rm{s}$ $~~~(5-year)$ \\
Galactic Accelerators & probe PeV accelerators up to 300 TeV & $E_{res}\sim$ 30-40\% ;  $F_{100\rm{TeV}}<10^{-13}$~erg/cm$^{2}.\rm{s}$   \\
PWNe \& $\gamma$-ray Halos & detect+resolve HGPS sources to 2$^{\circ}$ extension & angular resolution $\sim$ 0.2$^{\circ}$ \\
Diffuse Emission & detect Fermi Bubbles beyond 1 TeV  & excellent CR background rejection  \\
Dark Matter & probe full WIMP mass range to 100 TeV  & access to the GC and Halo   \\
Cosmic Rays & mass-resolved dipole anisotropy  & muon-counting capability   \\
\bottomrule
\end{tabular*}
\begin{tablenotes}
\item[$^1$] See \url{https://arxiv.org/pdf/1902.08429.pdf} for a detailed description of the full Science Case for SWGO.
\item[$^2$] See \url{https://arxiv.org/pdf/1907.07737.pdf} for a recent discussion of the proposed Observatory design plans.
\end{tablenotes}
\end{table*}
\end{center}

\paragraph{Dark Matter}
Among the fundamental physics topics to be covered by SWGO, constraining the nature of Dark Matter is certainly the principal science driver. In particular, given its geographical location, SWGO is in a unique position among current and planned facilities to search for DM signals from the direction of the Galactic Center, from where we expect to derive the best constraints, covering the entire WIMP mass range up to the highest energies of 100 TeV~\citep{2019JCAP...12..061V}.

In comparison for example with IACTs, the large field-of-view of SWGO will enable robust searches for the extended emission from decaying DM particles from the direction of the Galactic halo, as well as a rigorous quantification of any (diffuse) signal background. Above the 10 TeV range, where performance peaks, such backgrounds are expected to become small, and very good flux sensitivity is expected to be reached, down to few $\times$ 10$^{-14}$ erg/cm$^{2}$.s around 100 TeV.   

\paragraph{Cosmic rays}
SWGO will also be able to study directly cosmic rays with energies in the TeV -- PeV range. One of the fronts in which SWGO could play a difference in this energy range is on the study of the spectrum and composition of CRs. By efficiently measuring both the electronic and muonic components of the air showers at each WCD unity, we aim to reach a sensitivity enough to separate light and heavy mass groups, with a mass resolution goal roughly of, or better than $\Delta A/A\sim0.8$.

SWGO should also be able to play a crucial role by filling an important latitude gap for the study of cosmic-ray anisotropy, complementing and connecting the picture currently provided by Northern Hemisphere instruments and IceCube in the South Pole~\citep{2019ApJ...871...96A}. Here, we aim not only to extend current studies to smaller scale anisotropy, better constraining structures with $l \leq 15$ of the spherical harmonic expansion (or $\sim 10^{\circ}$ in angular scale), but also to probe the dipole swing for various (ideal four, $A = 1, 4, 14, 56$) CR species. 

\subsection{A wide-field observatory for gamma-ray transients}

Thanks to its proposed low-energy detection threshold, its wide field-of-view and continuous observational duty cycle, and its very large effective area, SWGO is uniquely placed to play a pivotal role in the field of transient astronomy and fill-up a missing niche in the global network of multi-messenger astronomy. In particular, it's Southern Hemisphere location provides the opportunity for synergies with CTA, both as a monitoring instrument and a trigger facility for high-energy transients. SWGO will be able to probe the entire range of variability timescales present in the gamma-ray sky (see Figure 4), from the extreme, sub-minute flaring activity of GRBs, to AGN and galactic transient sources which may sustain high emission states for  timescales of several days to weeks.

Among the possible transient science cases targeted by SWGO, GRBs and AGNs are the central ones, first of all because of their serendipitous character, requiring all-sky monitoring capabilities, and for their strong links to multi-messenger events -- either neutrinos~\citep{2017ATel10817....1M} or gravitational waves~\citep{2018JCAP...05..056P}.

In the case of GRBs, SWGO should be able not only to detect bright, nearby events early on, triggering follow-ups from CTA and other MWL instruments at the $\sim$ 1ks timescale, but it will also provide the opportunity to probe their VHE emission  during the prompt or early afterglow phases, virtually inaccessible to IACTs. This is a fundamental operational requirement to study e.g., short-GRBs, resulting from NS-NS merger events and linked to GW signals.

By pushing the observational threshold significantly below 300 GeV, with a good sensitivty $\sim 10^{-11}$ erg.cm$^2$.s, SWGO will open up the observational window for AGN monitoring at VHEs to sensitivity levels much beyond what is currently achievable~\citep{2017ApJ...843..116A}. In this way, regular outbursts ($\sim$ hour timescales) from a number of flaring blazars are expected to be detectable~\citep{2020MNRAS.497.3142L}. 

\subsection{Derived Performance goals}

\begin{figure}[t]
	\centerline{\includegraphics[width=106mm,height=14pc]{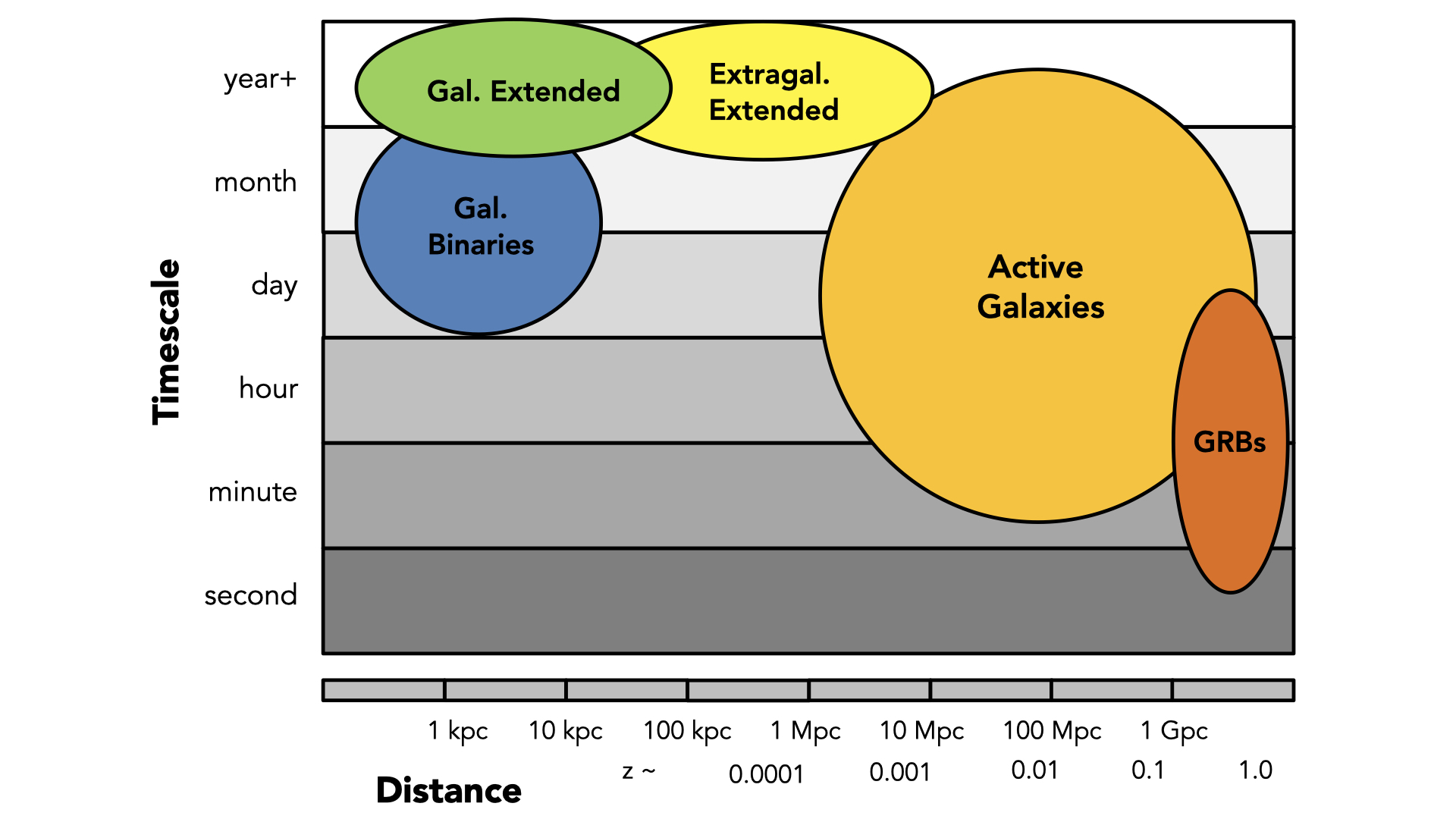}}
	\caption{VHE transient source classes and their variability timescales. SWGO will provide the necessary complement to observation of the VHE transient sky. Credits: Jim Hinton. \label{fig4}}
\end{figure}

The set of core science cases outlined above implies a number of performance requirements which will drive the observatory design during the current R\&D phase. A list of the crucial performance goals isT:
\begin{itemize}
\item A dense array core and excellent gamma-hadron separation capability to achieve low-energy detection threshold of c. 100 GeV (e.g.,~\cite{2018APh....99...34A}). 
\item An extended sparse array to achieve peak point source sensitivity at around 100 TeV.
\item Muon ID and counting capability at individual WCD unitis for cosmic-ray studies and improved background suppression.
\item Improved angular ($\sim$ 0.2$^\circ$) and energy ($\sim$ 30-40\%) resolutions throughout the core energy range.  
\end{itemize}

\section{Candidate Sites and Next Steps}

The R\&D phase of SWGO is set to last until 2022, after which a science-driven proposal for the observatory is to be presented. 

Beyond the definition of the science programme and the array design, the Collaboration is currently investigating the potential sites for installation of SWGO. The call for site proposals is currently open and a final shortlist of candidate sites is due next year, with a final decision expected for late 2021. 

Among the current list of sites under consideration are:

\begin{description}
\item[Argentina] Alto Tocomar (4450 m asl) and Cerro Vacar (4800 m asl). 
\item[Bolivia] Chacaltaya Plateau (4700 m asl). 
\item[Chile] Pajonales (4400 m asl) and Pampa la Bola (4700 m asl) 
\item[Peru] Imata \& Imata Lake (4450 m asl) and Laguna Sibinacocha (4900 m asl)
\end{description}



\nocite{*}
\bibliography{Wiley-ASNA}%

\end{document}